\documentclass[sigconf]{acmart}

\usepackage{setspace}
\usepackage{chronology}
\usepackage{lineno,hyperref}
\usepackage{pxfonts}
\usepackage{xcolor}
\usepackage{listings}
\usepackage[english]{babel}
\usepackage[utf8]{inputenc}
\usepackage[T1]{fontenc}
\usepackage{type1ec}
\usepackage{amssymb}
\usepackage{graphicx}
\usepackage{parcolumns}
\usepackage{color}
\usepackage{amsfonts}
\usepackage{colortbl}
\usepackage{booktabs}
\usepackage{amsmath}
\usepackage{tabularx}
\usepackage{textcomp}
\usepackage{soul}
\usepackage{footmisc}
\usepackage{changepage}
\usepackage[all]{nowidow}
\usepackage{enumerate}
\usepackage{multicol}
\usepackage{multirow}
\usepackage{balance}
\usepackage{array} 
\usepackage{varwidth}
\usepackage{scalefnt}
\usepackage{marginnote}
\usepackage{tcolorbox}
\usepackage[caption=false]{subfig}

\newcommand{\lenghtFigure}[0]{0.21}

\begin{document}

\title{Why We Engage in FLOSS: Answers from Core Developers} 

\author{Jailton Coelho, Marco Tulio Valente}
\affiliation{Federal University of Minas Gerais, Brazil}
\email{{jailtoncoelho,mtov}@dcc.ufmg.br}

\author{Luciana L. Silva}
\affiliation{Federal Institute of Minas Gerais, Brazil}
\email{luciana.lourdes.silva@ifmg.edu.br}

\author{André Hora}
\affiliation{Federal University of Mato Grosso do Sul, Brazil}
\email{hora@facom.ufms.br}

\copyrightyear{2018} 
\acmYear{2018} 
\setcopyright{acmcopyright}
\acmConference[CHASE'18]{CHASE'18:IEEE/ACM 11th International Workshop on Cooperative and Human Aspects of Software }{May 27, 2018}{Gothenburg, Sweden}
\acmBooktitle{CHASE'18: CHASE'18:IEEE/ACM 11th International Workshop on Cooperative and Human Aspects of Software , May 27, 2018, Gothenburg, Sweden}
\acmPrice{15.00}
\acmDOI{10.1145/3195836.3195848}
\acmISBN{978-1-4503-5725-8/18/05}

\begin{abstract}
The maintenance and evolution of Free/Libre Open Source Software (FLOSS) projects demand the constant attraction of core developers. In this paper,  we report the results of a survey with 52 developers, who recently became core contributors of popular GitHub projects. We reveal their motivations to assume a key role in FLOSS projects (e.g.,~improving the projects because they are also using it), the project characteristics that most helped in their engagement process (e.g.,~a friendly community), and the barriers faced by the surveyed core developers (e.g.,~lack of time of the project leaders). We also compare our results with related studies about others kinds of open source contributors (casual, one-time, and newcomers).
\end{abstract} 


\maketitle
	
\section{Introduction}
\label{sec:introduction}

Free/Libre and Open Source Software (FLOSS) projects have an increasing impact on our daily lives. For example, many companies depend nowadays on open source operating systems, databases, and web servers to run their basic operations. Similarly, most commercial software produced today depend on a variety of open source libraries and frameworks. However, there is a growing concern on the long term sustainability of FLOSS projects~\cite{nadia2016roads, hata2015characteristics}. For example, in a recent study, Avelino et al.~\cite{avelino2016novel} looked at a sample of 133 popular GitHub projects and concluded that nearly two-thirds depend on just one or two developers to survive~\cite{avelino2016novel}. For this reason, FLOSS projects must continuously attract new {\em core developers} to mitigate the risks of failing.

Core developers are the ones responsible for the design, implementation, and maintenance of the most important features in a project. They are also responsible to manage the project and to plan and drive its evolution~\cite{mockus2002two, joblin2017classifying, robles2009evolution}. By contrast, peripheral contributors are those who occasionally contribute to the projects, mostly by fixing bugs and implementing minor features~\cite{ye2003toward,pinto2016more,steinmacher2016overcoming}. Usually, core developers represent just a small fraction of the project contributors. For example, {\sc d3/d3}---a very popular JavaScript visualization library, with over 73K stars on GitHub---has 121 contributors. However, the system is maintained and evolved by just one core developer~\cite{avelino2016novel}.

Since core developers are the heart and brain of FLOSS projects, we report in this paper a survey with 52 developers who, in the last year, contributed to popular GitHub systems to the point of becoming core developers in these projects. By surveying these developers, our goal is to reveal their motivations for joining an open source project. We also asked them about the project characteristics that most helped in this process and about the main barriers they faced. The survey results can help FLOSS developers to improve some of the management practices followed in their projects, aiming to possibly expand the base of core developers.

We make the following contributions in this paper:

\begin{itemize}

\item We provide a list of motivations that led recent core developers to contribute to open source projects. We found that 60\% of the survey participants contribute because they are also using the projects.\\[-0.25cm]

\item We reveal a list of project characteristics and practices that helped recent core contributors to join a FLOSS project. We found they are most attracted by non-technical characteristics, especially the ones related to a friendly and available FLOSS community.\\[-0.25cm]

\item We provide a list of the main barriers faced by recent core contributors when joining a FLOSS project. We found that non-technical barriers are the most relevant impediment they face to contribute, as the lack of time of the project leaders.

\end{itemize}

\begin{figure*}[!ht]
  \centering
\subfloat[ref1][Age]{\includegraphics[width=\lenghtFigure  \textwidth]{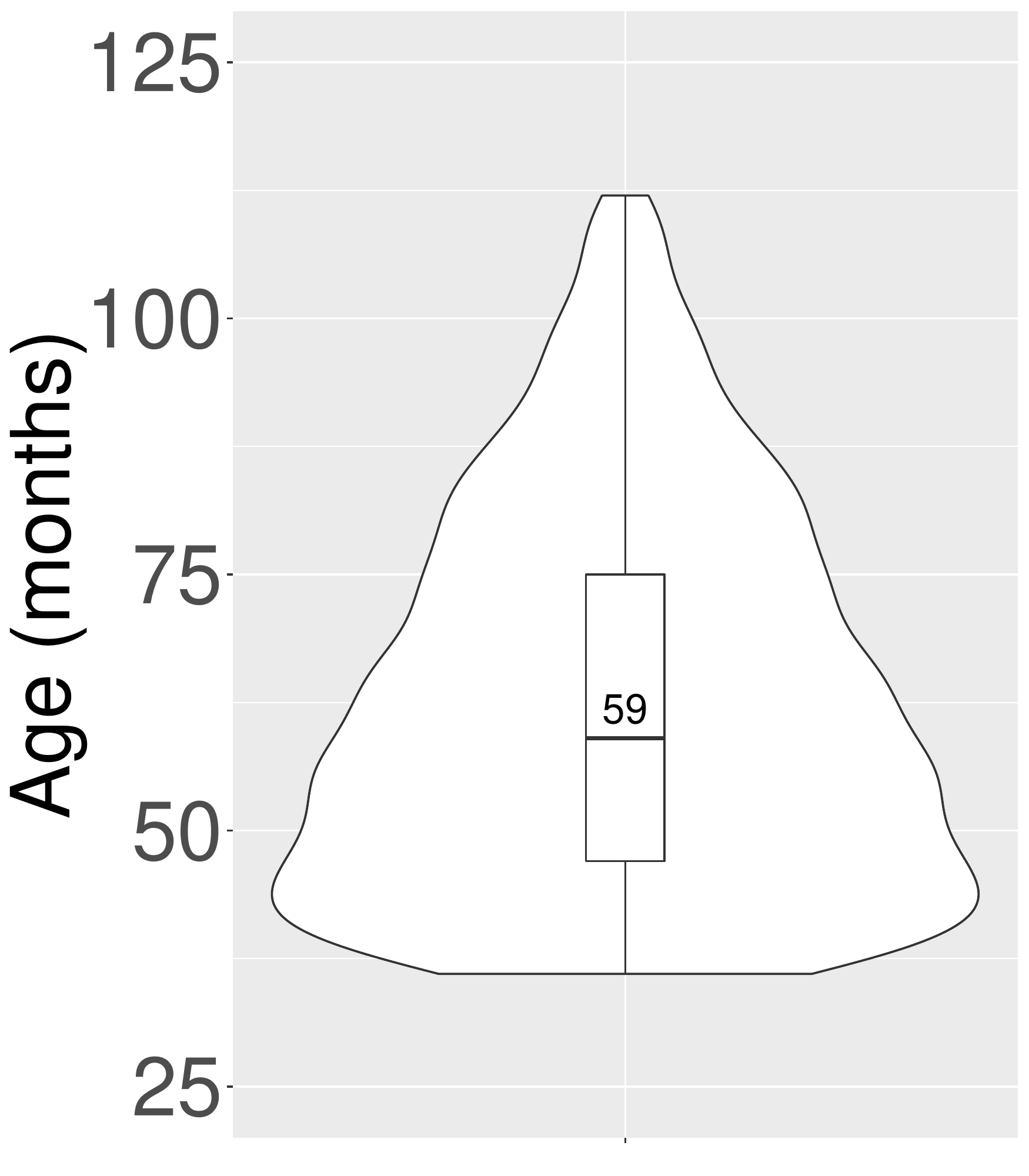}}
\quad
\subfloat[ref2][Contributors]{\includegraphics[width=\lenghtFigure \textwidth]{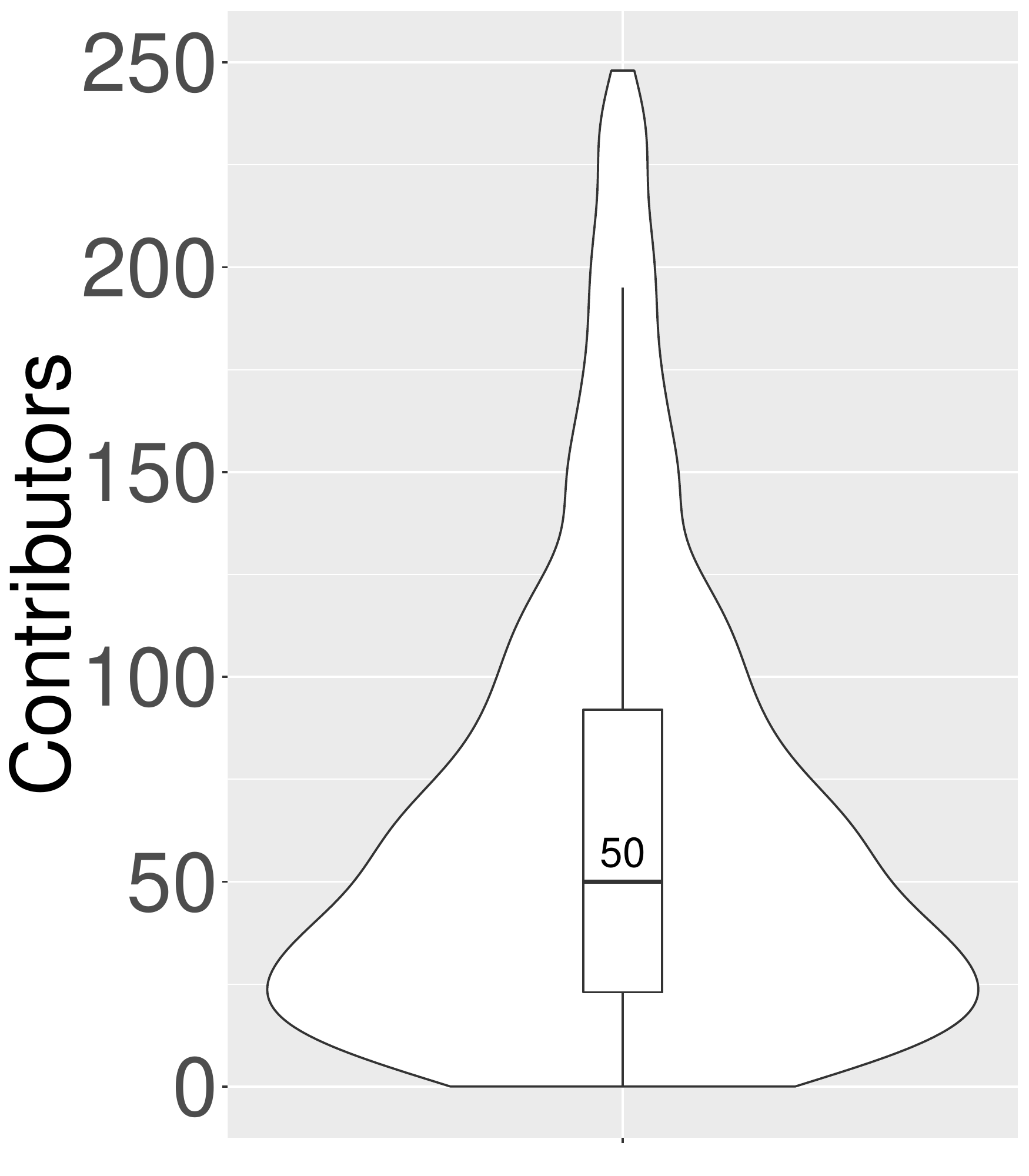}}
\quad
\subfloat[ref3][Commits]{\includegraphics[width=\lenghtFigure \textwidth]{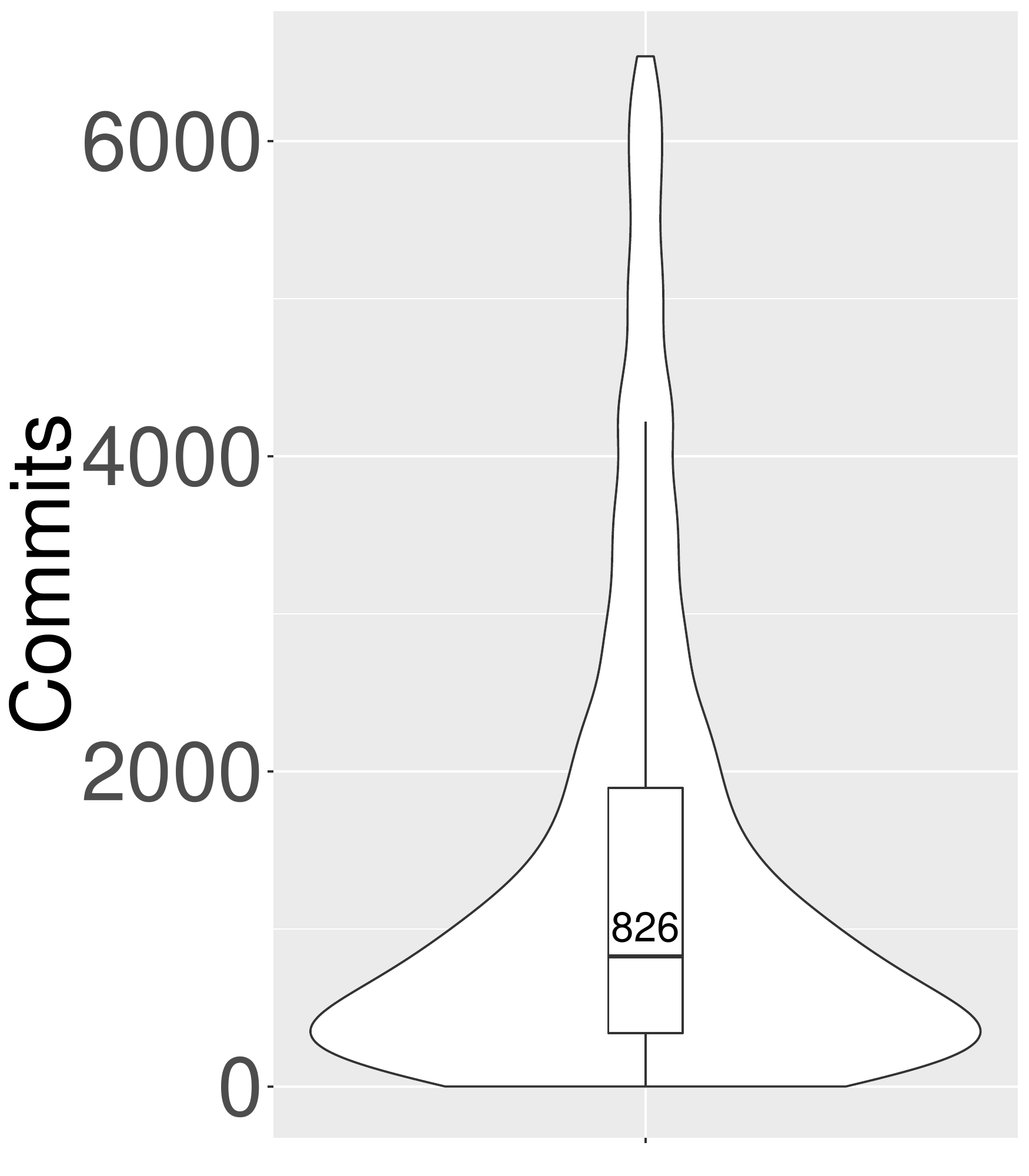}}
\quad
\subfloat[ref4][Stars]{\includegraphics[width=\lenghtFigure \textwidth]{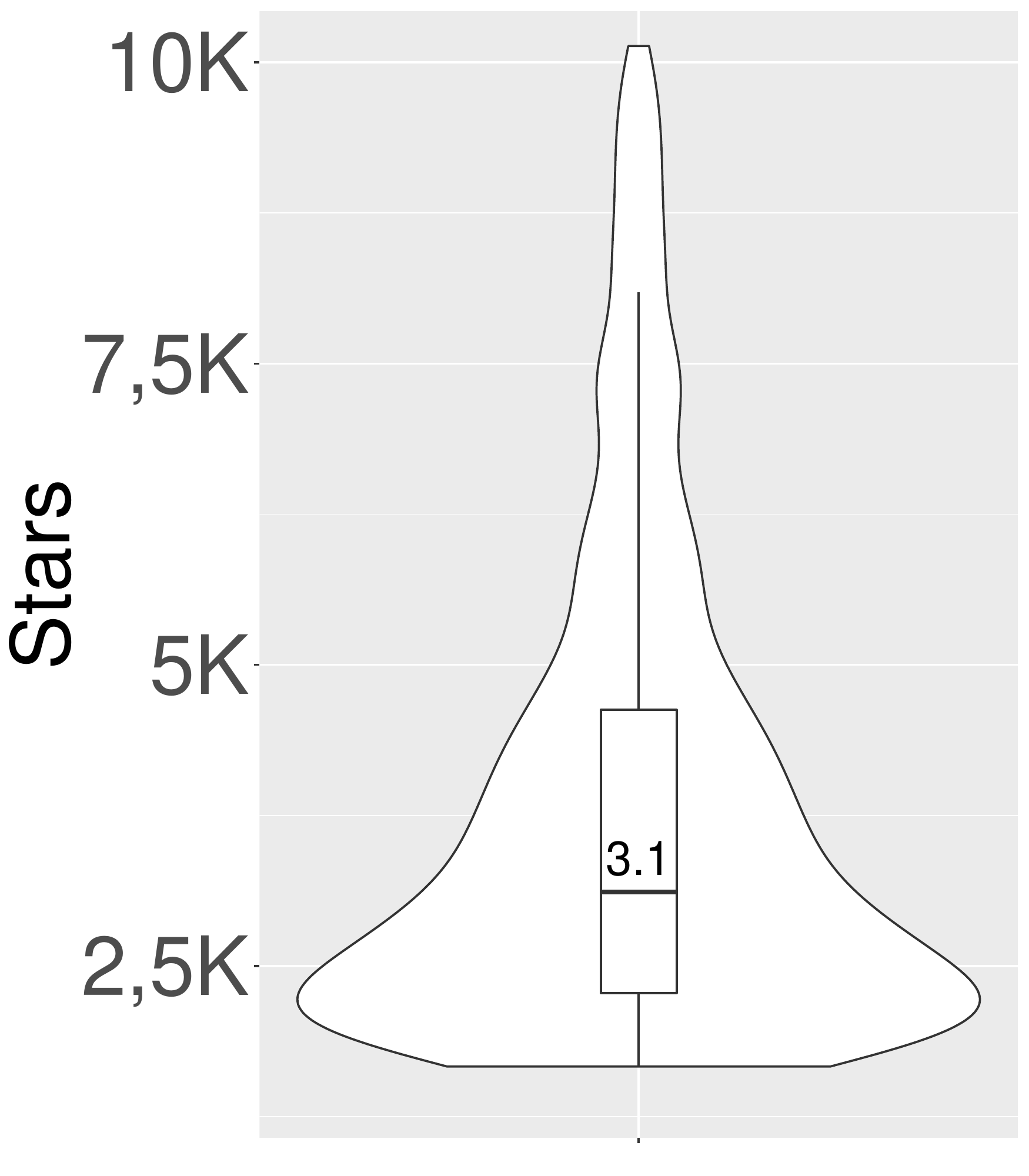}}
\caption{Distribution of the (a) age, (b) contributors, (c) commits, and (d) stars of the selected projects, without outliers.}
\label{dataset_violinplot_projects_characteristics} 
\end{figure*}

We organize the remainder of the paper as follows. Section~\ref{sec:study-design} presents the study design, how we selected the studied projects and the heuristic we used to identify core developers. Section~\ref{sec:survey-results} discusses the main findings of the survey. Section~\ref{sec:analysis-by-Project-Categories} presents a segmented analysis of the survey answers. Section~\ref{sec:threats} discuss threats to validity and Section~\ref{sec:related_studies} presents related work. Section~\ref{sec:implications} presents the main implications of our study, including implications to practitioners and researchers. Finally, Section~\ref{sec:conclusion} concludes the paper.

\section{Study Design}
\label{sec:study-design}

We start by considering the top-5,000 most popular GitHub projects, ranked by number of stars. Stars are similar to {\em likes} in popular social networks and therefore are a common measure of the popularity of GitHub projects~\cite{borges2016icsme}. Then, we apply four strategies to discard projects from this initial selection, as follows:

\begin{enumerate}
\item \textit{Non-Software Projects}: We discarded 61 repositories that are not software projects, including books (e.g., {\sc vhf/free-programming-books} and {\sc getify/You-Dont-Know-JS}) and awesome-lists (e.g., {\sc sindresorhus/awesome}). To remove these projects we relied on their GitHub topics. Specifically, we discarded projects with the topics {\em book} or {\em awesome-list}.\footnote{This step represents just a first attempt to remove non-software repositories; step (2) is also used to this purpose.}

\item \textit{Projects with no lines of code in a set of programming languages}: First, we used the tool {\sc AlDanial/cloc}\footnote{https://github.com/AlDanial/cloc} to compute the size of the projects, in lines of code (LOC). We configured this tool to only consider code in the top-100 most popular programming languages in the TIOBE list.\footnote{https://www.tiobe.com/tiobe-index} As a result, we discarded 397 projects, which are implemented in languages like HTML, CSS, and Markdown (i.e.,~in non-programming languages). For these projects, the size in LOC (counting only source code implemented in major programming languages) is equal to zero. As examples, we removed the following projects: {\sc github/gitignore} (which is a collection of textual .gitignore templates), {\sc jlevy/the-art-of-command-line} (a selection of notes and tips on using Linux command-line tools), and {\sc necolas/normalize.css} (a collection of HTML element and attribute style-normalization).

\item \textit{Inactive projects}: We are interested in projects under active development. Therefore, we discarded 830 repositories without commits in the last six months.
\item \textit{Non-mature projects}: Our central goal is to survey recent core developers of mature FLOSS projects. Particularly, it is important the projects have a minimal age in order to provide enough development time to compute new core developers. For this reason, we discarded 1,450 repositories with less than three years. 
\end{enumerate}

We ended up with 2,262 open source systems, including well-known projects, as {\sc facebook/react}, {\sc angular/angular}, and {\sc rails/rails}. Figure~\ref{dataset_violinplot_projects_characteristics} shows violin plots with the distribution of age (in months), number of contributors, number of commits, and number of stars of the selected projects, without  considering outliers. The median measures are 59 months, 50 contributors, 826 commits, and 3.1K stars, respectively. 1,256 projects (55\%) are owned by organizations and 1,006 repositories (45\%) by individual users. These projects are mainly implemented in JavaScript (696 projects, 31\%), followed by Ruby (232 projects, 10\%), and Python (230 projects, 10\%).

\begin{figure*}[!t]
  \centering
\subfloat[ref1][Core team contributions]
{
	\includegraphics[width=\lenghtFigure \textwidth]{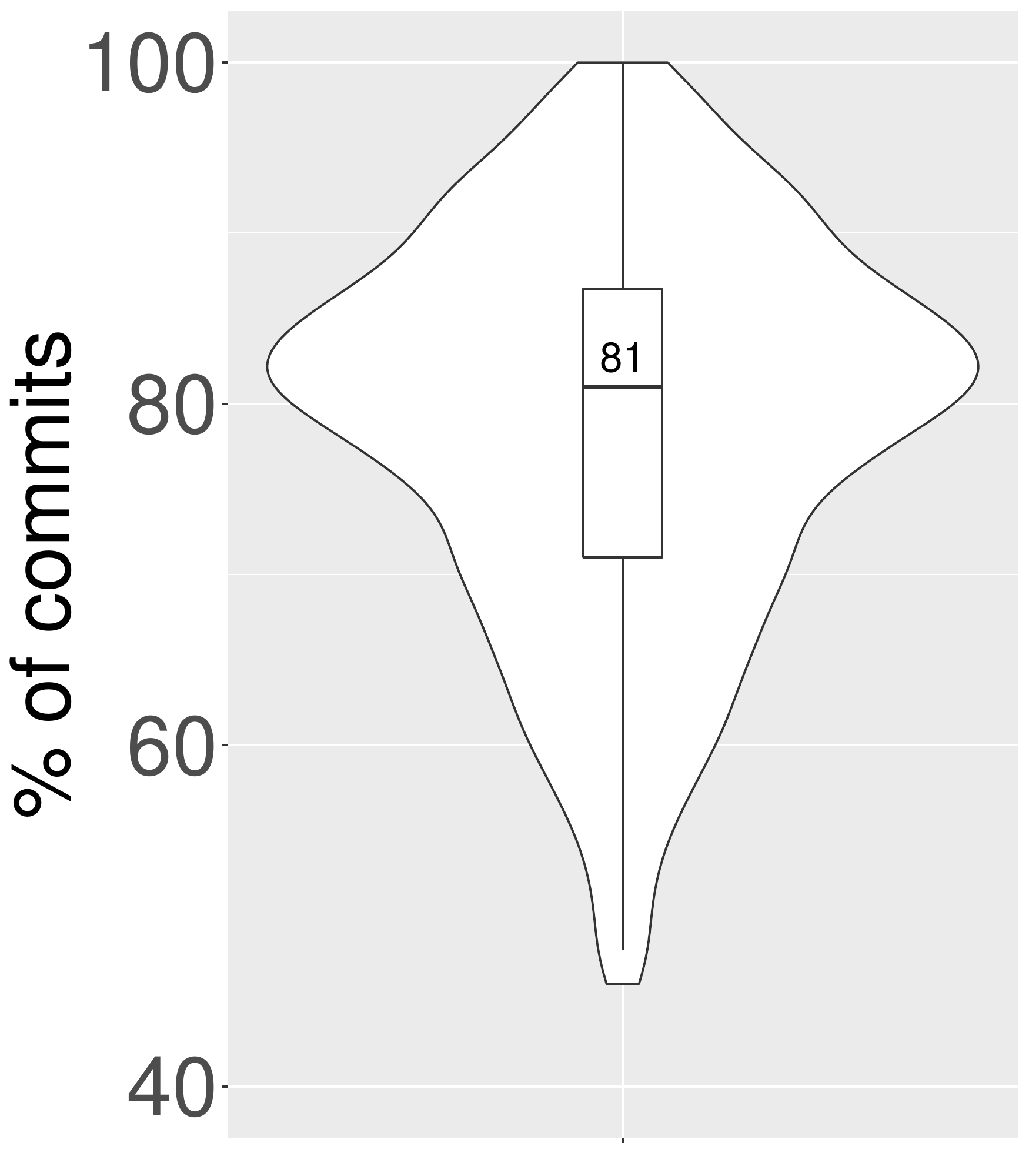}
}
\quad
\subfloat[ref2][Core team size]
{
	\includegraphics[width=\lenghtFigure \textwidth]{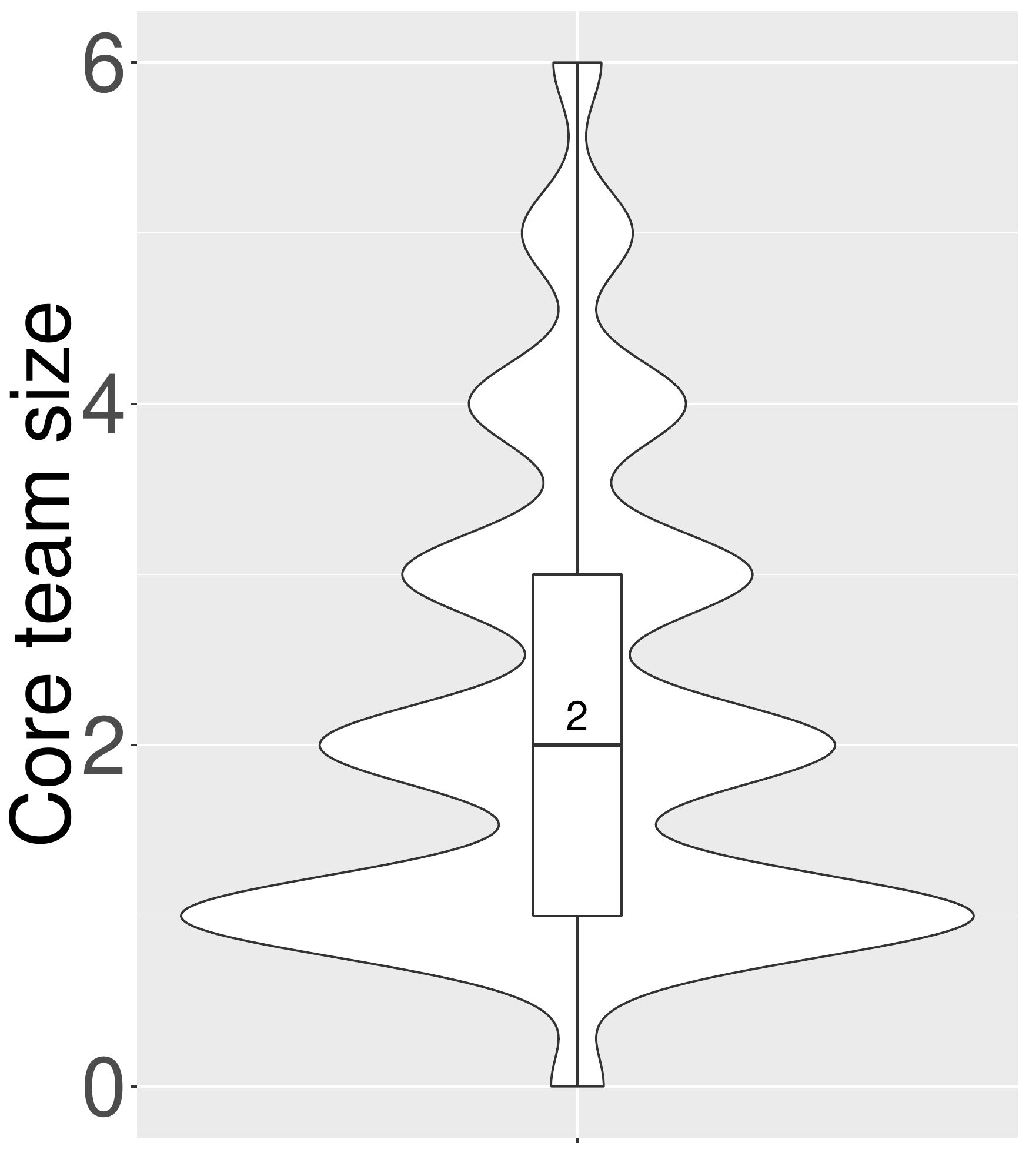}
}
\quad
\subfloat[ref3][Core developer contributions]
{
	\includegraphics[width=\lenghtFigure \textwidth]{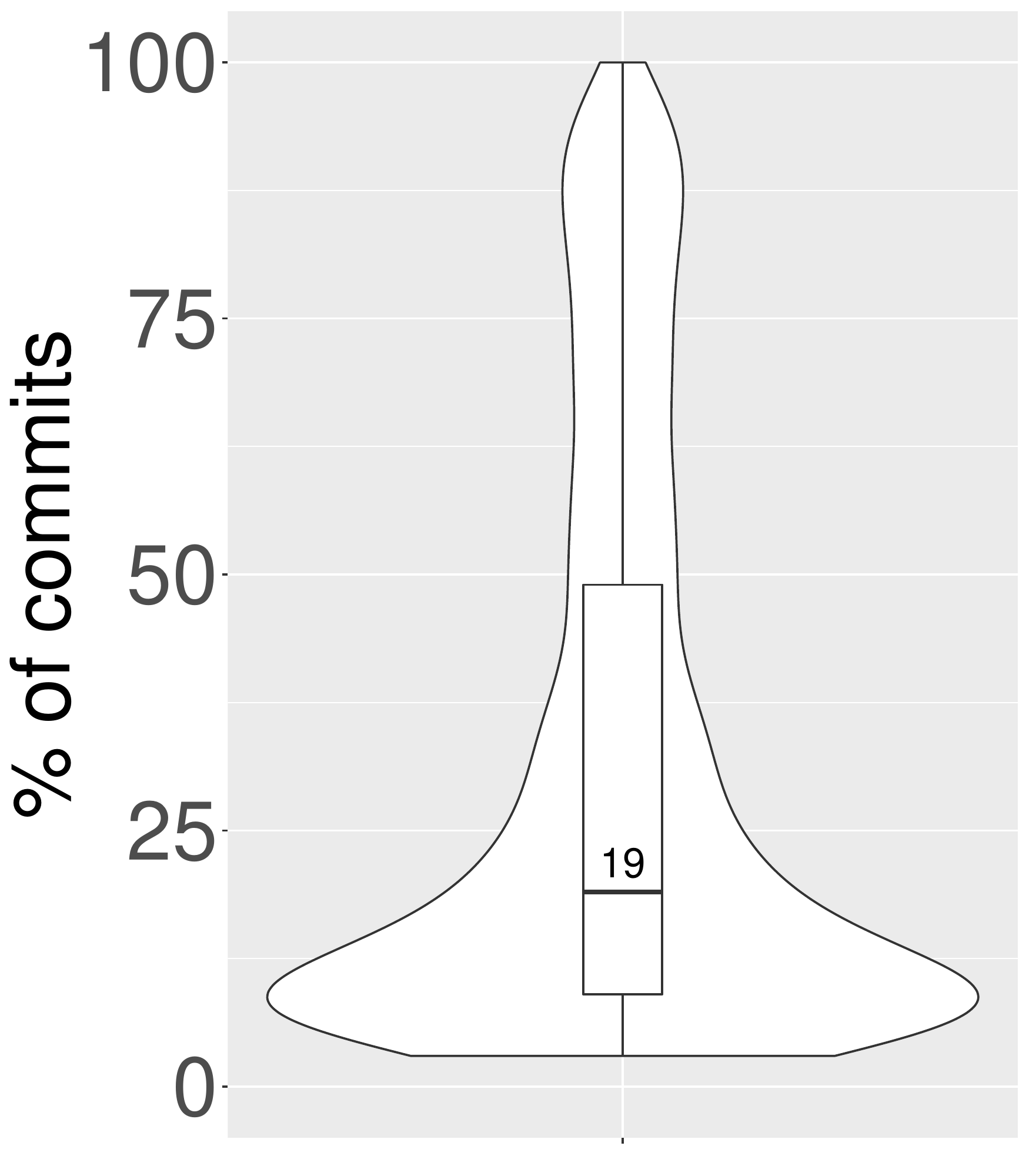} 
}
\caption{(a) Total percentage of commits by the selected core teams, (b) number of core developers per project, and (c) percentage of commits by the selected core developers. Outliers are omitted in these plots.}
\label{percent_of_contributions}
\end{figure*}

\subsection{Core Developer Identification}

To identify the core developers of each project, we use a Commit-Based Heuristic, which is commonly adopted in other studies~\cite{mockus2002two, robles2009evolution, dinh2005freebsd, koch2002effort}. This heuristic is centered on the number of commits by the project contributors, which usually follows a heavy-tailed distribution~\cite{mockus2002two,koch2002effort}, i.e., a minority of developers accounts for most contributions. According to this heuristic, the core team are those who produce 80\% of the overall amount of commits in a project. However, as usually defined, this heuristic accepts developers with few contributions, regarding the total number of commits. For this reason, we customized the heuristic after some initial experiments to require core developers to have at least 5\% of the total number of commits; candidates who have fewer commits are excluded. For example, to achieve 80\% of the commits in {\sc moment/moment}, the core team initially identified by the heuristic consists of 41 contributors. However, 38 contributors have less than 5\% of the overall amount of commits. Thus, only three developers are classified as \textit{core} by our customized heuristic. These developers represent 35\%, 25\% and 7\% of the project's commits, respectively. In favor of using this second threshold, the literature reports that even in complex projects, the core team is no bigger than 10-15 developers~\cite{mockus2002two}.

Despite the adoption of this second threshold, we observe in Figure~\ref{percent_of_contributions}a that the median percentage of commits by the selected core teams is 81\%. Figure~\ref{percent_of_contributions}b shows the core team size per project considering the minimal threshold of 5\%. We can see that more than half of the selected projects have only one or two core developers. Finally, as presented in Figure~\ref{percent_of_contributions}c, the median percentage of commits by the selected core developers is 19\%, in contrast to 0.5\% using the original strategy.

\begin{figure}[!h]
\centering
\includegraphics[width=8cm]{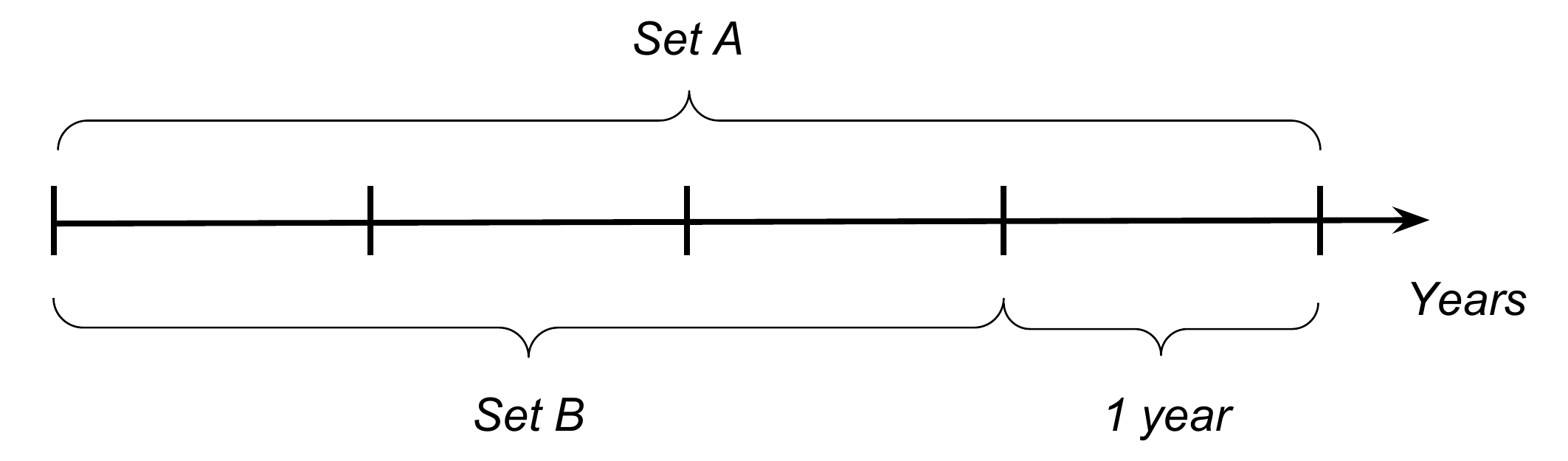}
\caption{Set A= core developers computed considering the complete commit history; Set B= core developers computed considering the commits until the year before the study; {\em New Core Developers} = {\em Set A} - {\em Set B}}
\label{fig:set-core-devs}
\end{figure}

Finally, we follow three steps to select developers who became core contributors in the last year of each project (see an illustration in Figure~\ref{fig:set-core-devs}):
(a) we apply the proposed heuristic on all commits of the project (set A);
(b) we remove the last year of commits and recalculate the core team (set B); and
(c) the selected set of {\em core developers} is formed by developers in the set A, but who are not in set B. In other words, this group includes developers who entered in the core team in the last year. We ended up with a list of 380 core developers, distributed over 331 projects.

\subsection{Survey Design}
\label{subsec:survey_design}

To some extent, our survey can be seen as a firehouse study, i.e.,~one that is conducted right after the event of interest has happened~\cite{fse2016-why-we-refactor, brito2018saner}.
Essentially, we surveyed {\em recent} core developers, to reveal their motivations to engage in FLOSS projects and the main barriers they faced during this process. After removing the core developers who do not have a public email address on GitHub, we obtained a list of 151 potential survey participants. We sent an email to these participants with two parts. First, we include the developer's name and data on his/her percentage of commits in the project. Then, the second part includes three open-ended questions about his/her contributions to the project: (1) What motivated you to contribute to this project? (2) What project characteristics and practices helped you to contribute? (3) What were the main barriers you faced to contribute?  

We received 52 answers (covering distinct projects), which corresponds to a response rate of 34\% (and a confidence interval of 11.04 for a confidence level of 95\%). Finally, we use Thematic Analysis~\cite{cruzes2011recommended} to interpret the survey answers. This technique is used for identifying and recording {\em themes} (i.e., patterns) in textual documents. Thematic Analysis consists of: (1) identifying themes from the answers, (2) reviewing the themes to find patterns for merging, and (3) defining and naming the final themes. The initial theme identification and merge steps were performed independently by the first two authors of this paper. Then, we had several meetings to resolve conflicts and define the final themes. In the first question, both authors suggested semantically equivalent themes for 32 answers (62\%). These themes were then rephrased and standardized to compose the final theme set. As the remaining 20 answers had divergent themes, they were discussed by both authors to reach a consensus. For the last two questions, an initial agreement was reached in 36 (69\%) and 38 (73\%) answers, respectively.

\section{Survey Results}
\label{sec:survey-results}

The presentation and discussion of the survey results are organized around the survey questions. To preserve the respondents’ anonymity, we use labels D1 to D52 to identify them. Furthermore, when quoting their answers we replace mentions to GitHub repositories, owners, and organizations by {\em[Project-Name]}, {\em[Project-Owner]}, and {\em[Organization-Name]}, respectively. 
This is important because some answers include sensitive comments about developers or organizations. It is also important to note that a question could have received two or more themes during the thematic analysis process.

\subsection{Motivations}

In the next paragraphs, we present the reasons that emerged for the first survey question ({\em What motivated you to contribute?})
We discuss each reason and also give examples of answers.

{\noindent \textbf {To improve the project because I am using it:}}
According to 31 new core developers, they increased their contributions primarily to fulfill their own needs. 
As examples, we have the following answers:\\[-.2cm]

\noindent{\em I started using it, I ran into minor issues or opportunities to improve, or things that were blocking me from making progress. Since it was an open source project, I was able to contribute improvements and make the project better for my needs, and everyone else's.} (D50)\\[-.25cm]

\noindent{\em First, I was a very active user of this project at the time. However, I felt that this software could be better. I believed I had enough experience to contribute, so I stepped in.} (D15)\\[-.25cm]

\noindent{\em I was using [Project-Name] for my startup in our internal dashboards and I needed a couple of features.} (D43)\\[-.2cm]

{\noindent \textbf {To have a volunteer work:}} 10 developers answered they contributed to take part in an open source community. As examples, we have the following answers:\\[-.2cm]

\noindent{\em I'm also in love with the idea of people sharing tools for free in order to help build a better world and promote scientific development and improving people's lives.} (D48)\\[-.25cm]

\noindent{\em I think that the fact that I was helping a lot of people, immediate feedback, motivated me to contribute more at the difficult time.} (D15)\\[-.2cm]

{\noindent \textbf {I have interest or expertise on the project domain:}} According to seven respondents, they were motivated by their interest or expertise on the project domain or programming language. As examples, we have these answers:\\[-.2cm]

\noindent{\em I've always had an interest in optimizing things, which I definitely did a lot of in this case.} (D06)\\[-.25cm]

\noindent{\em I'm well acquainted with the Ruby open source world...} (D10)\\[-.2cm]

{\noindent \textbf {I am a paid developer:}} Five new core developers mentioned they were paid to contribute, as in the following answer:\\[-.2cm]

\noindent{\em To be honest I'm paid for contributing to [Project-Name]...} (D23)\\[-.25cm]

{\noindent \textbf {To contribute to a widely used or relevant project:}} According to four developers, they were motivated by the fact the project is widely used or supported by well-known organizations. As examples, we have the following answer:\\[-.2cm]

\noindent{\em Working on a project as large as [Project-Name], and knowing that any work I contribute may be used by thousands of developers, was a pretty good motivator.} (D06)\\[-.2cm]

The remaining motivations are as follows: {\em because I know the maintainer} (3 answers), {\em to improve my programming skills} (2 answers), {\em to improve my CV} (2 answers), {\em because the project has a nice design and implementation} (1 answer), and {\em to train developers to contribute to FLOSS} (1 answer). 

Table~\ref{tab:RQ-One} summarizes the motivations reported by the participants for the first question. Among the 10 motivations mentioned by the developers, only two can be viewed as technical ones (e.g.,~{\em because I have interest or expertise on the project domain} and {\em because the project has a nice design and implementation}). The other motivations are non-technical and related to the interests of the developers or the project environment.

\begin{table}[!t]
    \centering
    \caption{What motivated you to contribute?}    
    \begin{tabular}{ l r l }
        \toprule
        {\bf Motivations}	& {\bf Dev.}	\\ 
        \midrule
        To improve the project because I am using it				& 	31 	\\
        To have a volunteer work									& 	10 	\\
        I have interest or expertise on the project domain			&	 7	\\
        I am a paid developer										&  	 5	\\
        To contribute to a widely used or relevant project			&  	 4	\\
        Other motivations ($\leq$ 3 answers each)					&  	 9	\\
        \bottomrule
    \end{tabular}
    \label{tab:RQ-One}
\end{table}

\subsection{Project Characteristics and Practices}

In the next paragraphs, we present the themes that emerged for the second survey question ({\em What project characteristics and practices helped you to contribute?}). We describe each reason and also provide examples of answers.\\[-.3cm]

{\noindent \textbf {Friendly community:}} According to 13 developers, they decided to increase their contributions due to the friendly community of project maintainers, who helped with issues and provided detailed feedback when revising pull requests. As examples, we have the following answers:\\[-.2cm]

\noindent{\em The main thing that helped me contribute was the friendliness of maintainers and the instructions they've left in the issues they answered.} (D48)\\[-.25cm]

\noindent{\em The [Project-Name] community gave very detailed feedback during pull requests (sometimes quite strict feedback!) which I found really helpful, and learned a lot about Git in the process.} (D27)\\[-.2cm]

{\noindent \textbf {Availability of the project leaders:}} According to 11 developers, the availability of the project leaders helped them to contribute, as in the following example:\\[-.2cm]

\noindent{\em In order for people to become contributors, in any kind of open source project, the most important thing is communication and availability of the owner/maintainer.} (D07)\\[-.2cm]

{\noindent \textbf {Unit tests:}} According to 9 respondents, the presence of unit tests helped them to increase the number of contributions. As example, we have the following answer:\\[-.2cm]

\noindent{\em Unit tests also helped a lot, allowed me to make changes freely with the comfort that I most likely haven't broken anything.} (D25)\\[-.2cm]

{\noindent \textbf {Documentation:}} This characteristic, as indicated by eight developers, refers to a clear and complete documentation. As examples, we have:\\[-.2cm]

\noindent{\em Extended documentation that helped to keep an idea of what it was all about: which things belongs to the project and which do not.} (D26)\\[-.25cm]

\noindent{\em Documentation for the whole code, especially documentation for setting up development environments of the project, I would really have struggled without that.} (D25)\\[-.2cm]

Table~\ref{tab:RQ-Two} summarizes the answers for the second question. 
In addition to the previously mentioned characteristics, we received answers citing {\em well-structured design} (4 answers),
{\em code review} (3 answers), {\em continuous integration} (3 answers), {\em programming language} (3 answers), {\em open source license} (3 answers), 
{\em small project} (3 answers), {\em coding guidelines} (2 answers), {\em clear code} (2 answers), {\em contribution guidelines} (2 answers), {\em financial support by private company} (1 answer), {\em large scale tests} (1 answer), and {\em small number of core developers} (1 answer).
In Table~\ref{tab:RQ-Two}, we also provide a classification of the developers answers in two major groups: technical and non-technical characteristics.

\begin{table}[!t]
	\centering
	\caption{What project characteristics/practices helped you?}
	\begin{tabular}{@{} l l r l@{} }
		\toprule
		{\bf Type} & {\bf Characteristics/Practices}			& {\bf Dev.}	\\ 
         \midrule
		\multirow{12}{*}{Technical} 
        & Unit tests								& 9		\\
		& Documentation								& 8		\\
		& Well-structured design					& 4		\\
		& Code review								& 3		\\
		& Continuous integration					& 3		\\
		& Programming language						& 3		\\
		& Open source license						& 3		\\
		& Small project								& 3		\\
		& Coding guidelines							& 2		\\		
		& Clear code								& 2		\\
		& Contribution guidelines					& 2		\\
		& Other technical characteristics			& 8		\\
		\midrule
		\multirow{5}{*}{Non-Technical}  
        & Friendly community						&  13	\\ 
        & Availability of the project leaders	 	&  11	\\ 
        & Financial support by a company			&  	1	\\
        & Open and meritocratic culture				&	1	\\
        & Small number of core developers			&  	1	\\       
		\bottomrule
	\end{tabular}
	\label{tab:RQ-Two}
\end{table}

\subsection{Barriers}

In this section, we present and discuss the themes that emerged for the third survey question ({\em What were the main barriers you faced to contribute?}). \\[-.3cm]

{\noindent \textbf {Lack of time of the project leaders:}} According to eight developers, the main barrier was the absence of the project leaders. As examples, we have the following answers:\\[-.2cm]

\noindent{\em Sometimes there were very slow replies to Issues/PRs as there were very few project leaders who could merge them.} (D20)\\[-.25cm]

\noindent{\em The original developer basically stopped working on it years ago. Many of us were still using the plugin, but bug reports and pull requests built up for years without attention.} (D38)\\[-.2cm]

{\noindent \textbf {Large and complex project:}} Seven developers answered that project complexity and size were the main barriers they faced to increase their contributions, as in the example:\\[-.2cm]

\noindent{\em The project as a whole is complex and requires specialized knowledge or skill sets that I don't always have.} (D45)\\[-.2cm]

{\noindent \textbf {Non-clear, complex or buggy codebase:}} According to five respondents, the main barrier concerns a non-clear, complex or buggy codebase. As example, we have the following answer:\\[-.2cm]

\noindent{\em The code was plagued with race conditions, code smells, bad practices and ugly workarounds. This made it very hard for me to quickly make changes.} (D41)\\[-.2cm]

{\noindent \textbf {Inappropriate design or architecture:}} This barrier, mentioned by four respondents, refers to inappropriate design or architecture. As example, we have the following answer:\\[-.2cm]

\noindent{\em Less than optimal project structure or release structures.} (D07)\\[-.2cm]

\begin{figure*}[!t]
  \centering
\subfloat[ref1][Project characteristics and practices]
{
	\includegraphics[width=0.48 \textwidth]{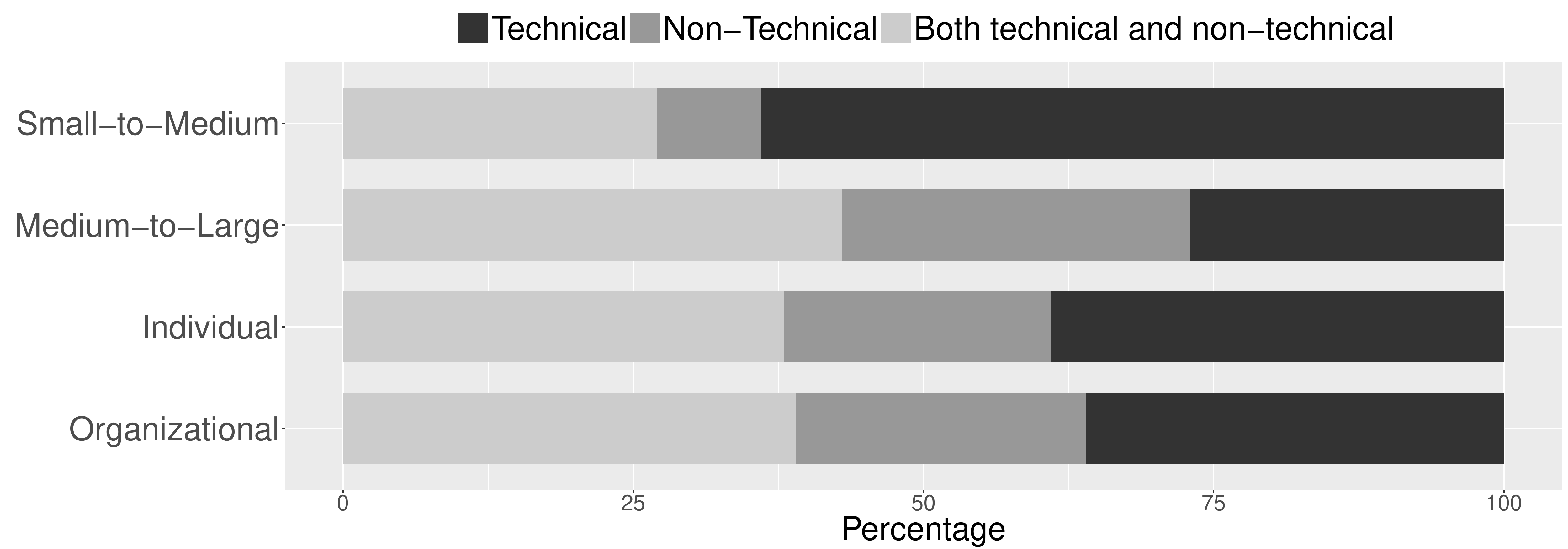}
}
\quad
\subfloat[ref2][Barriers]
{
	\includegraphics[width=0.48 \textwidth]{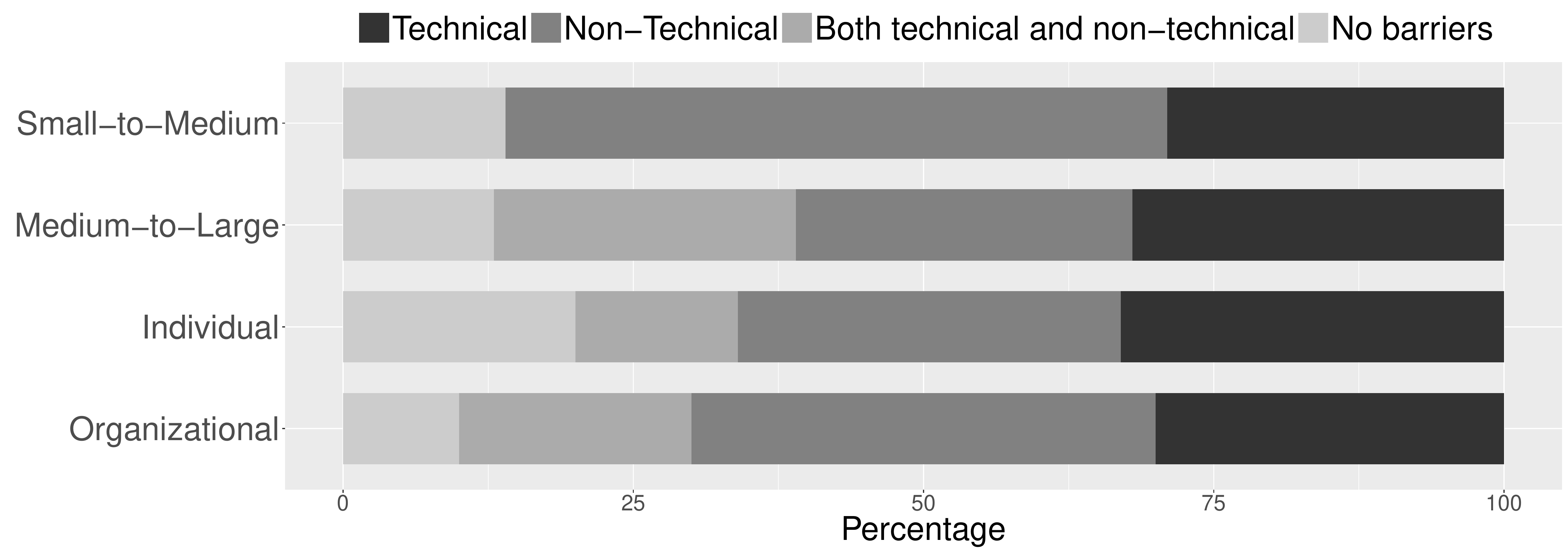}
}
\caption{Results grouped by project categories (Small-to-Medium, Medium-to-Large, Individual, and Organizational)}
\label{fig:proj-grouped}
\end{figure*}

Table~\ref{tab:RQ-Three} summarizes the responses for the third question.  In addition to the previously mentioned barriers, we received answers citing {\em inexperience of the own contributor} (3 answers), {\em lack of time of the own contributor} (3 answers), {\em lack or incompleted documentation} (3 answers), {\em programming language} (3 answers), {\em lack of tests} (3 answers), and {\em conflicts among developers} (3 answers). Furthermore, {\em English language}, {\em decisions must be approved by a committee}, {\em old coding styles}, {\em hostile attitudes}, {\em lack of build tools}, {\em project requires specialized knowledge} are other mentioned barriers, all of them with a single answer. Finally, six (11\%) participants answered they faced no barriers at all. As we can see, there in this case a balance between technical and non-technical barriers, which received 33 and 27 answers, respectively.

\begin{table}[!t]
	\centering
	\caption{What were the barriers you faced to contribute?}
    \small
	\begin{tabular}{ @{}l l r l@{} }
		\toprule
		{\bf Type}	&	{\bf Barriers}					& {\bf Dev.}	\\ 
		\midrule
        \multirow{7}{*}{Technical} 
		& Large and complex project						& 7		\\
		& Non-clear, complex or buggy codebase			& 5		\\
		& Inappropriate design or architecture			& 4		\\			
		& Lack or incompleted documentation				& 3		\\
		& Programming language							& 3		\\	
		& Lack of tests									& 3		\\
        & Other technical barriers						& 8		\\
        \midrule
		\multirow{7}{*}{Non-Technical} 
        & Lack of time of the project leaders			& 8		\\
        & Lack of time of the own contributor	 		& 4		\\
        & Conflicts among developers					& 3		\\
        & Inexperience of the own contributor	 		& 3		\\	
        & Hostile attitude								& 1		\\
        & Unpaid work									& 1		\\
        & Other non-technical barriers					& 7		\\        
		\bottomrule
	\end{tabular}
	\label{tab:RQ-Three}
\end{table}

\section{Analysis by Project Categories}
\label{sec:analysis-by-Project-Categories}

In this section, we provide results grouped by the following categories of projects: {\em small-to-medium} vs {\em medium-to-large} projects and {\em individual} vs {\em organizational} projects.\\[-.2cm]

{\noindent \textbf {Project Categories:}} We classify the 52 projects according to their size, considering the distribution of LOC of the 2,262 projects. The projects in the first and second quartiles are classified as {\em Small-to-Medium} projects (LOC $\leq$ 4,894); the ones in the third and fourth quartiles are named {\em Medium-to-Large} projects (LOC > 4,894). We ended up with a list of 17 {\em Small-to-Medium} and 35 {\em Medium-to-Large} projects.

We also group the 52 projects considering the type of the account on GitHub: 18 projects are developed using individual accounts (e.g.,~{\sc javan/whenever}) and 34 projects using an organizational account (e.g.,~{\sc google/guava}).\\[-.2cm]


{\noindent \textbf {Results:}} Figure~\ref{fig:proj-grouped}a shows the results for project characteristics and practices. The figure shows the percentage of technical, non-technical and both technical and non-technical characteristics. For example, developers contributed to individual projects exclusively due to their technical (39\%), non-technical (23\%), and both technical and non-technical characteristics (38\%). According to the results in  Figure~\ref{fig:proj-grouped}a, technical characteristics are the most important factor in small-to-medium projects (64\%).
By contrast, they are exclusively responsible to the engagement of core developers in only 27\% of the medium-to-large projects. In these projects, most answers include a combination of technical and non-technical factors (43\%). Finally, there is no major difference in the results for individual and organizational projects. For example, technical factors are the only factors responsible by the engagement of core developers in 39\% of the individual projects and 36\% of the organizational ones.

\begin{table*}[!t]
	\centering
	\caption{Comparison of our findings with related studies.}
	\footnotesize
	\begin{tabularx}{\textwidth}{ @{}l | X | X | X | X@{}}
		\toprule
		\multicolumn{1}{c|}{\bf Topic}  					& \multicolumn{1}{c|}{\bf Our study}  																					& \multicolumn{1}{c|}{\bf Lee et~al.~\cite{leeunderstanding}}  														& \multicolumn{1}{c|}{\bf Pinto et~al.~\cite{pinto2016more}} 												 	& \multicolumn{1}{c}{\bf Steinmacher et~al.~\cite{steinmacher2016overcoming}}\\ 	 
		\midrule
		\begin{tabular}[c]{@{}l@{}} Contributors \end{tabular}			& \begin{tabular}[c]{@{}X@{}} Core developers \end{tabular}																		& \begin{tabular}[c]{@{}X@{}} One-Time code Contributors \end{tabular}													& \begin{tabular}[c]{@{}X@{}} Casual contributors \end{tabular}													&  \begin{tabular}[c]{@{}X@{}} Newcomers \end{tabular}		\\
		\midrule		
		\begin{tabular}[c]{@{}l@{}} Motivations\end{tabular}			& \begin{tabular}[c]{@{}X@{}}\\[-.2cm]To improve the project because I am using it\\\\[-.2cm] To have a volunteer work\\\\[-.2cm] I have interest or expertise on the project domain\end{tabular} 	& \begin{tabular}[c]{@{}X@{}}\\[-.2cm]To fix bugs\\\\\\[-.2cm] To give back to the community\\\\\\[-.2cm]I am a paid developer \end{tabular}				& \begin{tabular}[c]{@{}X@{}}\\[-.2cm]To fix bugs\\\\\\[-.2cm] To improve documentation\\\\\\[-.2cm] To add new features \end{tabular}				& \multicolumn{1}{c}{-}						\\
		\midrule
		\begin{tabular}[c]{@{}l@{}} Project\\characteristics \end{tabular} 	& \begin{tabular}[c]{@{}X@{}}\\[-.25cm]Friendly community\\\\[-.2cm]Availability of the project leaders\\\\[-.2cm]Unit tests  \end{tabular}								& \begin{tabular}[c]{@{}X@{}}\\[-.25cm]Skilled project members\\\\[-.2cm]Friendly project members\\\\[-.2cm]Helpful project members \end{tabular} 			& \multicolumn{1}{c}{-}	 																	& \multicolumn{1}{|c}{-}				\\
		\midrule
		\begin{tabular}[c]{@{}l@{}} Barriers \end{tabular} 			& \begin{tabular}[c]{@{}X@{}}\\[-.25cm]Lack of time of project leaders\\\\[-.2cm] Large and complex project\\\\[-.2cm] Unclear, complex or buggy code \end{tabular}					& \begin{tabular}[c]{@{}X@{}}\\[-.25cm]Lack of time of own contributor\\\\[-.2cm] Complex submission process\\\\[-.2cm] Complex project \end{tabular} 			& \begin{tabular}[c]{@{}X@{}} \\[-.25cm]Lack of time of own contributor\\\\[-.2cm] Limited skills or knowledge\\\\[-.2cm] Complex project \end{tabular}		&  \begin{tabular}[c]{@{}X@{}} \\[-.25cm]Technical barriers\\\\[-.2cm]  Lack of contribution guidelines\\\\[-.2cm] Lack of documentation \end{tabular}	\\
		\bottomrule
	\end{tabularx}
	\label{tab:related-work}
\end{table*}


Figure~\ref{fig:proj-grouped}b shows the breakdown results for the barriers faced by the surveyed developers. First, the percentage of projects presenting no barrier at all ranges from 10\% (organizational projects) to 20\% (individual projects). The most common barriers in small-to-medium projects are exclusively non-technical ones (57\%). In other words, developers contribute  to small-to-medium projects due their technical characteristics, but often face non-technical barriers. As example, we have this answer from a core developer about the  technical characteristics and practices of a small-to-medium project:
\\[-.2cm]

\noindent{\em Proper coding guidelines and documentations do help a lot. (D05) } \\[-.2cm]

But he also complained about non-technical barriers:\\[-.2cm]

\noindent{\em Not all contributors have a consistent and equivalent share of time to invest in the project. This sometimes stalls the progress \ldots (D05)}\\[-.2cm]

Regarding medium-to-large projects, we found a balance among technical barriers (32\%), non-technical barriers (29\%) and both types of barriers (26\%). As in the case of project characteristics, there is no major difference in the results for individual and organizational projects.

In summary, we found that core developers are engaged in small-to-medium projects mostly due to their technical characteristics (e.g.,~unit tests), but often face non-technical barriers (e.g.,~lack of time of the project leaders). In medium-to-large projects, the surveyed core developers increased their contributions due to a combination of both technical and non-technical characteristics; they also faced both technical and non-technical barriers. Finally, we found that there is no major difference between individual and organizational projects, regarding their characteristics and offered barriers.

\section{Threats To Validity}
\label{sec:threats}

The threats to validity of this work are as follows:\\[-.25cm]

\noindent{\bf External Validity:}  The dataset used in this study is restricted to popular open source projects on GitHub.
We acknowledge that there are popular projects in other platforms (e.g.,~Bitbucket and GitLab) or projects that have their own version control infrastructure.\\[-.2cm]

\noindent{\bf Internal Validity:} This threat relates to the themes denoting the survey answers. 
We acknowledge that the selection of these themes is to some extent subjective. 
For example, it is possible that different researchers reach a different set of motivations, practices and barriers, than the ones proposed in Section~\ref{sec:survey-results}. To mitigate this threat, the initial theme selection was performed independently by the first two authors of this paper. After this initial proposal, several meetings were performed to refine and improve the initial selection.\\[-.2cm]

\noindent{\bf Construct Validity:}  A construct validity threat relates to the commit-based heuristic for core developer identification. However, we decided to use a traditional heuristic to this purpose, widely used in other studies~\cite{dinh2005freebsd, koch2002effort, mockus2002two}. Furthermore, we customized this heuristic to exclude developers with few commits (less than 5\%  of the total number of commits). 

\section{Related Work}
\label{sec:related_studies}

In this section, we first compare our results with related studies which focused on three profiles of open source contributors:

\begin{itemize}
\item {\em One-Time Code Contributors (OTC)} are developers who have exactly one accepted patch. \citet{leeunderstanding} conduct a survey with OTCs to comprehend their impressions, motivations, and barriers, when contributing to FLOSS. 

\item {\em Casual Contributors} are those with few contributions (e.g., less than 2\% of the total number of commits) and who do not want to become active project members.\footnote{To clarify, OTCs have just one contribution, while casual contributions can have more than one contribution.}
\citet{pinto2016more} conduct surveys with (1) casual contributors to understand what motivates them to contribute and (2) with project maintainers to understand how they perceive casual contributions.

\item {\em Newcomers} are those contributors who attempted to conclude their first contribution to an open source project. \citet{steinmacher2016overcoming} elicit 58 barriers that may hinder newcomers onboarding to open source projects.

\end{itemize} 

Table~\ref{tab:related-work} contrasts our results with  the aforementioned studies. The most common motivation for OTCs and casual contributors is {\em bug fixing} because it can affect their work. In contrast, the most common motivation for core developers engagement, as revealed in our survey, is {\em improving the project because I am using it}. Therefore, their motivation include not only bug fixing tasks, but also adding new features.~\citet{leeunderstanding} investigate impressions that increase the chances of a potential developer to contribute to a project. The most common positive impression reported by OTCs is the presence of skilled, friendly, and helpful project members. Similarly, we found that core developers are also attracted by a {\em friendly community} and by the {\em availability of the project leaders}. However, the third characteristic cited by core developers is the presence of {\em unit tests}, while on the case of OTCs are helpful project members. The most common barrier faced by OTCs and casual contributors is {\em lack of time of the own contributor}. By contrast, only three core developers reported this fact as a main impediment. 

In a previous work~\cite{coelho2017why}, we conduct an investigation with maintainers of 104 open source projects that failed to understand the reasons of such failures. The most common reasons are projects that were usurped by competitors (27 projects),  obsolete projects (20 projects), lack of time of the main contributors (18 projects), and lack of interest of the main contributors (17 projects). \citet{robles2014floss} describe a curated dataset with data from over 2,000 FLOSS contributors. Among the collected data, this dataset includes the contributors motivations for joining FLOSS. However, these answers can be seen as general motivations; in our survey, we decided to ask specific developers (core developers) about their motivations for recently
joining well-defined open source projects. 

In a recent survey promoted by GitHub with thousands of open source developers, documentation was indicated as a pervasive problem when contributing to open source, according to 93\% of the respondents (see http://opensourcesurvey.org/2017). However, in our survey, restricted to core developers, documentation is mentioned as barrier by only three participants. 

\citet{nadia2016roads} reports on the risks and challenges to maintain open source projects. \citet{ye2003toward} describe a study to understand what motivates developers to engage in open source development. Other studies on open source have focused on how to attract and retain contributors~\citep{zhou2015will, canfora2012going, bosu2014peer}. \citet{gousios2015work,gousios2016work,gousios2014exploratory} provide insights on the pull-based development model as implemented in GitHub from the integrator and contributors’ perspective. \citet{mirhosseini2017can} investigate the use of pull request notifications in GitHub projects. \citet{jiang2016and} examine why and how developers fork repositories on GitHub. They found that developers fork repositories to submit pull requests, fix bugs, add new features, and keep copies. \citet{joblin2017classifying} categorized core and peripheral developers based on social and technical perspectives.

\section{Implications}
\label{sec:implications}

Our study has implications both to practitioners and researchers, as follows:\\[-0.2cm] 

\noindent{\bf Implications to Practitioners:} \underline{First}, core contributors should strive to provide an interesting and high-quality software product, which can attract a large base of users. Then, some of these users will decide to improve the product to fulfill their own needs. Finally, they will share the improvements with the project community, which can trigger a new cycle of improvements. \underline{Second}, two non-technical practices are important to engage core developers in open source projects: nurturing a friendly community and being always available. However, technical factors---specially, the availability of unit tests and documentation---are also important. \underline{Third}, the main barrier faced by new core contributors is also non-technical, the lack of time of project leaders, followed by two technical ones: project complexity and low quality code.\\[-0.2cm]

\noindent{\bf Implications to Researchers:} \underline{First}, open source projects are increasingly important elements of the digital infrastructure that supports our modern societies~\cite{nadia2016roads}. We also know that these projects depend on a small number of core developers~\cite{mockus2002two,avelino2016novel}. Thus, researchers should continue to investigate strategies to improve open source practices and communities. Particularly, our findings might contribute to current efforts to develop health and analytics models and tools to open source projects, as proposed for example by the CHAOSS\footnote{https://chaoss.community/} and SECOHealth~\cite{mens2017towards} projects. \underline{Second}, we also showed the importance of requiring a minimal percentage of commits when identifying core developers. When this threshold is not applied, the traditional heuristic can select core developers with very few commits, which are included just to reach the total amount of 80\% of the commits of a system.

\section{Conclusion}
\label{sec:conclusion}

In this paper, we reported the main reasons that led recent core developers to contribute to open source projects. We also reveal the most common project characteristics and practices that motivated them to engage in FLOSS and the barriers they faced to contribute. As future work, we plan to conduct interviews with selected project contributors, to validate and extend our findings. We are also working on a tool to assess the  ``health'' of FLOSS projects.

\section*{Acknowledgments}
Our research is supported by CAPES, FAPEMIG, and CNPq.

\bibliographystyle{ACM-Reference-Format}
\bibliography{icse2017}

\end{document}